\documentclass[sn-mathphys]{sn-jnl}
\jyear{2021}%

\theoremstyle{thmstyleone}

\theoremstyle{thmstyletwo}

\theoremstyle{thmstylethree}

\usepackage{natbib}
\usepackage{booktabs,longtable,array,ragged2e}
\usepackage{adjustbox,lipsum}

\begin{document}

\title[ExoSGAN and ExoACGAN]{ExoSGAN and ExoACGAN: Exoplanet Detection
using Adversarial Training Algorithms
}

\author*[1]{\fnm{Cicy K} \sur{Agnes}}\email{cicykagnes@ug.cusat.ac.in}

\author*[1]{\fnm{Akthar} \sur{Naveed V}}\email{akthercse2018@ug.cusat.ac.in}

\author[2]{\fnm{Anitha} \sur{Mary M O Chacko}}\email{anithamarychacko@cusat.ac.in}

\affil[1]{\orgdiv{Division of Computer Science and Engineering}, \orgname{Cochin University of Science and Technology}, \orgaddress{\street{Kalamassery}, \city{Kochi}, \postcode{682022}, \state{Kerala}, \country{India}}}

\affil[2]{\orgdiv{Assistant Professor, Division of Computer Science and Engineering}, \orgname{Cochin University of Science and Technology}, \orgaddress{\street{Kalamassery}, \city{Kochi}, \postcode{682022}, \state{Kerala}, \country{India}}}

\abstract{Exoplanet detection opens the door to the discovery
of new habitable worlds and helps us understand how
planets were formed. With the objective of finding earth-like
habitable planets, NASA launched Kepler space telescope and
its follow up mission K2. The advancement of observation
capabilities has increased the range of fresh data available
for research, and manually handling them is both time-consuming and difficult. Machine learning and deep learning techniques can greatly assist in lowering human efforts to process the vast array of data produced by the modern
instruments of these exoplanet programs in an economical
and unbiased manner. However, care should be taken to detect
all the exoplanets precisely while simultaneously minimizing
the misclassification of non-exoplanet stars. In this paper, we
utilize two variations of generative adversarial networks, namely
semi-supervised generative adversarial networks and auxiliary
classifier generative adversarial networks, to detect transiting
exoplanets in K2 data. We find that the usage of these models
can be helpful for the classification of stars with exoplanets. Both
of our techniques are able to categorize the light curves with a
recall and precision of 1.00 on the test data. Our semi-supervised
technique is beneficial to solve the cumbersome task of creating
a labeled dataset.}


\maketitle
\section{Introduction} \label{sec1}
Exoplanets, also known as extrasolar planets, are planets that orbit stars out side our solar system. For centuries humans have questioned if additional solar systems exist among the billions of stars in the universe. Despite numerous dubious claims of exoplanet discovery, Wolszczan and Frail are credited with discovering the first verified exoplanet in 1992. Exoplanet discovery is still in its early phases. The hunt for exoplanets helps us comprehend planet formation and the discovery of Earth-like habitable planets. It aids in the collection of statistical data and information about exoplanet atmospheres and compositions, as well as their host stars. It can also help us understand how the solar system formed.
In total, $\sim$4884 exoplanets have been identified as of December 2021\footnote{\url{https://exoplanetarchive.ipac.caltech.edu/docs/counts_detail.html}}. The standard exoplanet detection techniques include direct imaging, Doppler Spectroscopy , astrometry, microlensing, and the transit method \citep{2014}. Most of the known planets have been discovered by the transit method. If we detect periodic drops in flux intensities of the star when a planet passes in front of it, we can confirm that a planet transits around the star. In this paper we use the transit method of analysing light curves. However, analysing the flux curve to find them is a very tedious task because vast amounts of data are produced from the observatories that are often noisy.  

The search of exoplanets took a step forward with the help of observatories such as Kepler \citep{2016RPPh...79c6901B}, the CoRoT spacecraft \citep{10.1111/j.1365-2966.2011.19970.x}, the Transiting Exoplanet Survey Satellite (TESS) \citep{2015JATIS...1a4003R}, and others \citep{2006MNRAS.373..799C,refId0,refId01}. Although the Kepler mission concluded a few years ago and the data is well-systemised and publicly available, it is still far from being completely put to use. These data can provide insights that could pave the way for future discoveries in the coming decades. Usually professional teams manually inspect the light curves for possible planet candidates. They vote for each other to reach a final decision after concluding their study \citep{2018yCat..22390005C,2019yCat..51580025Y}. To properly and quickly  determine the presence of these planets without any manual efforts, it will be necessary to automatically and reliably assess the chance that individual candidates are, in fact, planets, regardless of low signal-to-noise ratios-the ratio of the expected dip of the transit to the predictable error of the observed average of the light intensities within the transit.
\begin{figure}[t]
\centering
\includegraphics[width=0.9\textwidth]{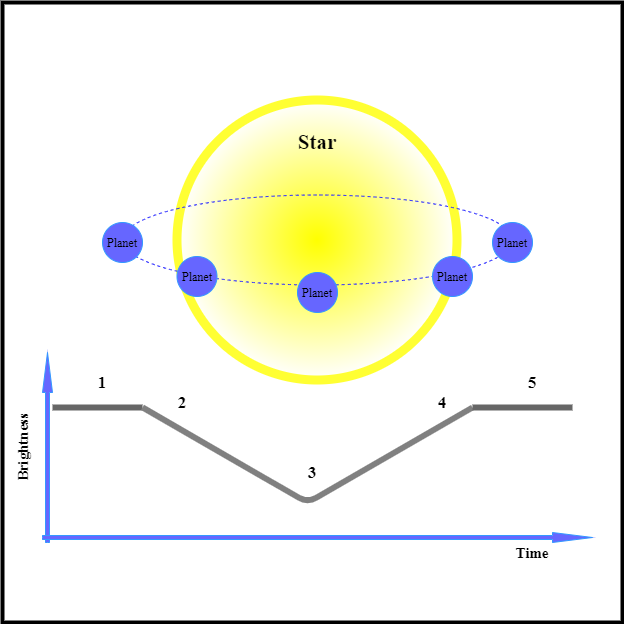}
\caption{Transit photometry: The flux intensity of the star varies at different point of time as the planet moves around the star. The presence of periodic dips in the intensity curves can confirm the presence of an exoplanet.}\label{fig:figure1}
\end{figure}
Furthermore, exoplanet classification is an example of an imbalanced binary classification problem. The issue with the data is that the available  number of stars with exoplanets is far lesser than the number of stars with no exoplanets. Fortunately, the advancements in machine learning help us to automate the complex computational tasks of learning and predicting patterns in light curves within a short time. In 2016, a robotic vetting program was developed to imitate the human analysis employed by some of the planet candidate catalogs in the Kepler pipeline \citep{2016A}. The Robovetter project was a decision tree meant to eliminate the false-positive ‘threshold crossing events’(TCEs) and to determine whether there are planets having a size similar to that of the earth present around any Sun-like stars. Likewise, in the past several other notable attempts such as the autovetter \citep{McCauliff_2015} were made to propound the applications of machine-learning methodologies in spotting exoplanets in  Kepler light curve data. 

Since the introduction of the Astronet, a deep learning model that produced great results in automatic vetting of Kepler TCEs \citep{2018} the advantages of using deep learning models for the classification of the light curves are more investigated.
Despite the fact that deep learning methods are computationally expensive, since they offer better results for complex problems, many researchers attempt to shift their attention from classical machine learning methods to deep learning methods. Also, as deep learning is a dynamic field that is constantly evolving, there are possibilities of new techniques other than the conventional models that might prove to be more efficient in exoplanet detection. 
Also, as astronomical projects include long-term strategies and are critical in educational, environmental and economical aspects, researches should adhere to accuracy and ethics. The biased research could cynically affect the scientific training, education and financing of future astronomers. In exoplanet detection we should not falsify or leave out any potential exoplanet candidates and fully harness the data available from the missions. 
Being motivated by these, we propose two deep learning models, Semi Supervised Generative Adversarial Networks (SGAN) and Auxiliary Classifier Generative Adversarial Networks (ACGAN), for the classification of exoplanets. We call these methods  ExoSGAN and ExoACGAN respectively. They produce comparable and sometimes better  classification results to the techniques so far used.

The paper is organized into six sections. Section \ref{sec2} includes relevant works in this field. Section \ref{sec3} contains an explanation of the methodologies utilized in this article. Section \ref{sec4} discusses the materials used,  data preprocessing prior to creating our model along with the architecture used to create the models. Section \ref{sec5} discusses the evaluation measures utilized and the findings obtained. The study closes with a section that summarises the findings and future directions in this research field.

\section{Related Works}\label{sec2}
A lot has happened in the realm of astrophysics as well as deep learning in the last 25 years. Many researchers are investigating the use of deep learning algorithms for exoplanet discovery. 
One of the widely used techniques for the detection of extrasolar planets include Box-fitting Least  Squares(BLS)\citep{2002A&A...391..369K}. The approach is based on the box-shape of recurrent light curves and takes into account instances with low signal-to-noise ratios. They use binning to deal with the massive number of observations. They point out that an appreciable identification of a planetary transit necessitates a signal-to-noise ratio of at least 6. Cases that appear to be a good fit are then manually assessed. However, the method is exposed to the risk of false-positive detections generated by random cosmic noise patterns.

A transit detection technique that makes use of the random forest algorithm is the signal detection using random forest algorithm (SIDRA) \citep{10.1093/mnras/stv2333}.  SIDRA was trained on a total of 5000 simulated samples comprising 1000 samples from each class namely constant stars, transiting light curves, variable stars, eclipsing binaries and microlensing light curves.  20,000 total samples from these different classes were tested where a success ratio of 91 percent is achieved for transits and 95-100 percent for eclipsing binaries, microlensing, variables and constant light curves. They recommended that SIDRA should only be used in conjunction with the BLS method for detecting transiting exoplanets since SIDRA had generated low results on transiting light curves.  

In 2017, Sturrock et al \cite{sturrock} conducted a study on the use of the machine learning methods namely, random forest (RF), k-nearest neighbour (KNN) and Support Vector Machine (SVM) in classifying cumulative KOI data (Kepler Objects of Interest). In terms of the fraction of observations identified as exoplanets, SVM did not give the desired prediction results. The RF model has the risk of overfitting the dataset. Inorder to reduce overfitting, measures such
\begin{table}
 \centering
\caption{Table summarizing the relevant works done in the field of exoplanet discoveries}
\label{table:table1}
\begin{adjustbox}{width=\textwidth,totalheight=0.95\textheight,keepaspectratio}

\begin{tabular}{p{3cm}p{4cm}p{4cm}}
   \toprule
    Year and author      & Methodology     &Results \\ \midrule
    
Mislis et al, 2015 \cite{10.1093/mnras/stv2333}                   
& Presented SIDRA, a random forest based algorithm for detection and classification of light curves. 
& Success ratio : 91 percent for transiting exoplanets
\\ \midrule
 
McCauliff et al, 2015 \cite{McCauliff_2015}            
& Presented the Autovetter project: a random forest model for classification of  exoplanets using Kepler light curves    
&Overall error rate - 5.85 percent   
\\  \midrule
        
Thompson et al, 2015 \cite{2015ApJ...812...46T} 
&Dimensionality reduction and k nearest neighbour technique were used to detect transit signals in Kepler light curves
&eliminates over 90 percent of the TCEs that do not have transiting exoplanet signals while keeping over 99 percent of the planet candidates and almost 99 percent of transit signals injected into the light curve signals 
\\ \midrule

Armstrong et al, 2016 \cite{10.1093/mnras/stw2881}  
&Employed Self organizing maps and random forests to rank transiting exoplanet candidates in Next generation transit survey (NGTS) data &AUC score - 97.6 percent 
\\ \midrule      

Sturrock et al, 2018 \cite{sturrock}                    
& Created a Random forest model to classify KOI data and and deployed it to public 
& Resulted in a cross validated Accuracy of  98 percent 
\\ \midrule
           
Shallue and Vanderburg, 2018 \cite{2018}  
&To identify transit signals of exoplanets, a CNN model named Astronet, trained on Kepler light curves, was proposed.
&AUC - 0.988,

Accuracy 96 percent on Kepler false positives.
and Accuracy 94.9 on kepler planet candidates.  
\\  \midrule

Dattilo et al, 2019 \cite{2019} 
&Developed Astronet K2 : a 1-d CNN model adapted from Astronet and applied it to K2 data
&Accuracy - 98 percent  
\\ \midrule

Malik et al, 2021 \cite{10.1093/mnras/stab3692} 
&Built a Gradient Boosted Tree(GBT) and used 10-fold cross validation for transit detection in Kepler and TESS(NASA’s Transiting Exoplanet Survey Satellite) data

&Kepler data: AUC- 0.948, Recall - 0.96, Precision - 0.82
 
TESS data: Recall - 0.82, Precision - 0.63
\\ \midrule
 
Yip et al, 2020 \cite{10.1007/978-3-030-46133-1_20}  
&used a GAN model to generate a suitable dataset for discovering exoplanets via the direct imaging technique and trained a CNN model for the exoplanet  classification task using the new dataset.  
&When signal-to-noise ratio(SNR) of the training set is 0.75, the accuracy on the test set (over different SNR) is 0.896 to 0.967
\\ \midrule

Priyadarshini and Puri, 2021 \cite{Priyadarshini2021} 
&Developed a CNN based ensemble model for K2 data by using RF, SVM, Decision trees and  Multilayer Perceptron as base model and CNN as metamodel 
&Accuracy - 99.62 percent \\ \midrule

Armstrong et al, 2020 \cite{10.1093/mnras/staa2498}  
&Demonstrated use of GPC (Gaussian Process Classifier) supported by other machine learning models for Kepler light curves  
&GPC model :  AUC - 0.999, Precision - 0.984, Recall - 0.995 \\ \bottomrule

\end{tabular} 
\end{adjustbox}
\end{table}
as StratifiedShuffleSplit, feature reduction, cross validation etc had to be done. It was found that random forest gave a cross-validated accuracy of 98 percent which is the best among the other three methods. Using Azure Container Instance and an API (application programming interface) the random forest classifier was made accessible to public.

In 2018, Shallue and Vanderburg \cite{2018} researched about identifying exoplanets with deep learning. This is a notable work in the field and is considered to be the state of the art method. They used NASA Exoplanet Archive: Autovetter Planet Candidate Catalog dataset and introduced a deep learning architecture called Astronet that utilizes convolutional neural networks(CNN). There are 3 separate input options to the network: global view, local view and both the global and local views. These input representations were created by folding the flattened light curves on the TCE period and binning them to produce a 1 dimensional vector. During training they augmented the training dataset by using random horizontal reflections. They used the google-vizier system to automatically tune the hyperparameters including the number of bins, the number of fully connected layers, number of convolutional layers, dropout probabilities etc. After model optimization they used model averaging on independent copies of the model with different parameter initializations to improve the performance by not making the model depend upon different regions of input space. The best convolutional model received AUC (Area Under the Curve) of 0.988 and accuracy of 0.960 on the test set. 

Astronet K2, a one-dimensional CNN with maxpooling, was trained in 2019 to discover planet candidates in K2 data \citep{2019}. This deep learning architecture is adapted from the model created by Shallue and Vanderburg \cite{2018}. It resulted in a 98 percent accuracy on the test set. EPIC 246151543 b and EPIC 246078672 b were discovered as genuine exoplanets by Astronet K2. They are both in between the size range of the Earth and Neptune. Looking at the precision and recall of Astronet K2, if the classification threshold is set such that the model offers a recall of 0.9 in the test set of Kepler data, the precision rate will only be 0.5. For Astronet K2 to estimate planetary occurrence rates in K2 data the model should be enhanced to produce a recall of 90 percent while simultaneously maintaining a precision of 95 percent because K2 data contains more false positive samples.

Later in 2020, 'TSfresh' library and 'lightbgm' tool was used in exoplanet detection by Malik et al\cite{10.1093/mnras/stab3692} . An ensemble of decision trees called gradient boosted trees (GBT) and 10-fold CV (cross validation) technique were used to create a model to classify light curves into planet candidates and false positives. Their prediction results in an AUC of 0.948, precision of 0.82 and recall of 0.96 in Kepler data. They provide comparable results to Shallue and Vanderburg \cite{2018} and also prove that their method is more efficient than the conventional BLS( box least squares fitting). However, the performance of this method was poor on the class imbalanced TESS data.

Yip et al \citep{10.1007/978-3-030-46133-1_20} used generative adversarial networks (GAN) to detect planets through direct imaging technique. GAN was used to create a suitable dataset which is further trained using convolutional neural networks classifier that can locate planets over a broad range of signal-to-noise ratios. 
 
Most of the studies related to using artificial intelligence in exoplanet detection utilize the random forest algorithm \citep{20191,10.1093/mnras/sty3146} and CNNs \citep{Ansdell_2018,2019MNRAS.488.5232C}. Other heuristic approaches to find extrasolar planets using stellar intensities include KNN \citep{2015ApJ...812...46T} and  self-organizing maps \citep{10.1093/mnras/stw2881}. In 2021 Priyadarshini and Puri \cite{Priyadarshini2021} used an ensemble-CNN model  on the Kepler light curves.   Different machine learning algorithms such as Decision Tree, Logistic Regression, MLP (Multilayer Perceptron), SVM( Support Vector Machines, CNN, random forest classifier and their proposed ensemble-CNN model were implemented and compared with each other. They used a stacked model that is trained similarly as k-fold validation. The decision tree, RF, SVM, and MLP models were the meta learners and CNN model was employed as base learner in their proposed work. The model was capable to produce an accuracy of 99.62 percent. However training ensemble models can be expensive and hard to interpret since we deal with light curves.

Furthermore, the use of semi supervised generative adversarial algorithm (SGAN) have been proved to be efficient in retrieving potential radio pulsar candidates \citep{20211}. The study indicates that SGAN outperforms standard supervised algorithms in real world classification. The best performing model in the study gives an overall F-score of 0.992. This model has been already incorporated into the HTRU-S Lowslat survey post-processing pipeline, and it has found eighteen additional pulsars.
   
Moreover, auxiliary classifier generative adversarial networks (ACGAN) have been useful in situations where imbalanced data is available for study. For instance, Wang et al \citep{8771369} developed a framework that gives improved and balanced performance for detecting cardiac arrhythmias using a data augmentation model and ACGAN. They also find that ACGAN based data augmentation framework gives better classification results while addressing imbalanced data.  Additionally, Sutedja et al \citep{bib2} suggested an ACGAN model to tackle an imbalanced classification issue of fraud detection in debit card transaction data. The study also compares the classification performance of the ACGAN model used with that of a CNN model-having similar architecture to the ACGAN discriminator model. They believe that because the ACGAN model produces better outcomes, it may also be used to solve unbalanced classification issues.

Table \ref{table:table1} summarises some of the main works done in the field so far. These studies reveal that exoplanet detection issue has been approached using standard artificial intelligence techniques but it remains a challenge that deserves more attention. Hardly any exploration has been done that reveals the use of novel deep learning techniques in astronomy. We capitalize the usage of the irrefutably appealing efficiency of semi-supervised generative adversarial networks and auxiliary classifier generative adversarial networks to tackle the issue. These are already proved to be useful in biomedical applications and other fields.

\section{Methods} \label{sec3}
\subsection{Generative Adverserial Networks}
Goodfellow et al (2014) \cite{goodfellow2014generative} introduced Generative Adversarial Networks in 2014. A generator G and a discriminator D are the two major components of this machine learning architecture. Both of them are playing a zero-sum game in which they are competing to deceive one other. The concept of the game can be summarized roughly as follows: The generator generates images and attempts to fool the discriminator into believing that the produced images are real. Given an image, the discriminator attempts to identify whether it is real or generated. The notion is that by playing this game repeatedly, both players will improve, which implies that the generator will learn to produce realistic images and the discriminator will learn to distinguish the fake from the genuine. 

G is a generative model which maps from latent space z to the data space X as $G(z;\theta_1)$ and tries to make the generative samples $G(z)$ as close as possible to the real data distribution $P_{data} (x)$. $P_{z}(z)$ is the noise distribution. D [given by $D(x;\theta_2)$] discriminates between the real samples of data(labeled 1) and the fake samples (labeled 0) generated from G and outputs $D(x) \in (0,1)$. 
The loss function of GAN can be derived from the loss function of binary cross entropy.
\begin{equation}\label{eqn1}
L(\hat{y},y)\,=\,[y\log\hat{y}\,+\,(1-y)\log(1-\hat{y})]
\end{equation}
where y is the original data and $\hat{y}$ is the reconstructed data. The labels from $P_{data} (x)$ is 1. $D(x)$ and $D(G(z))$ are the discriminator's probabilities that x is real and G(z) is real. Substituting them in equation \ref{eqn1} for real and fake samples, we get the two terms $\log(D(x))$ and $\log(1-D(G(z))$. Inorder to classify real and fake samples D tries to maximize the two terms. Generator’s task is to make $D(G(z))$ close to 1, therefore it should minimize the two terms above. Thereby, they follow a two player minmax game.
\begin{figure}[t]
\centering
\includegraphics[width=0.9\textwidth]{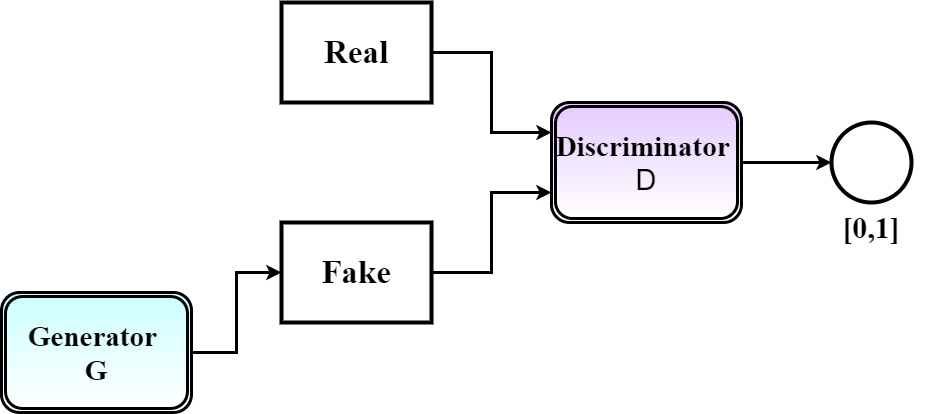}
\caption{
An overview of the GAN framework at a high level. The generator generates fake samples. Discriminator accepts false and real data as input and returns the likelihood that the provided input is real. 
}
\label{fig:figure2}
\end{figure}
The loss function of GAN can be expressed as follows, 
\begin{equation}\label{eqn2}
\begin{aligned}
\underset{G}{\min} \, \underset{D}{\max} \, V(D,G) \:= 
{} & \ E_{x \sim P_{data}(x)}[\,\log D(x)\,]\,	 +\,E_{z \sim P_{z}(z)}[\, \log(1-(D(G(z))\,]
\end{aligned}
\end{equation}

E represents expectation.

\subsection{Semi-Supervised Generative Adversarial Networks(SGAN)}
An extension to GAN was made by Odena (2016) \cite{odena2016semisupervised} called SGAN to learn a generative model and a classification job at the same time. Unlike the traditional GAN that uses sigmoid activation function to distinguish between real and fake samples, SGAN uses softmax function to yield N+1 outputs. These N+1 include the classes 1 to N and the fake class. Salimans et al (2016) \cite{salimans2016improved} presented a state of the art method for classification using SGAN on MNIST, CIFAR-10 and SVHN datasets at that time.
In our case SGAN should produce outputs-exoplanets, non exoplanets and fake samples.
Taking x as input, a standard supervised classifier produces the class probabilities (stars having exoplanets and stars having no exoplanets). We achieve semi-supervised learning here by providing the generated fake samples from G labeled y = N+1 to the D and categorising them as fake samples thereby introducing an additional dimension.
Let the model predictive distribution be $(P_{model}(y\vert x))$ as in case of a standard classifier where \(y\) is the label corresponding to \(x\). $P_{model}(y=N+1\vert x)$ gives the probability that x is fake. 
The loss function of SGAN consists of an unsupervised and a supervised component. 
\begin{equation}
L = L_{unsupervised}+L_{supervised}
\end{equation} 
\begin{figure}[t]
\centering
\includegraphics[width=0.9\textwidth]{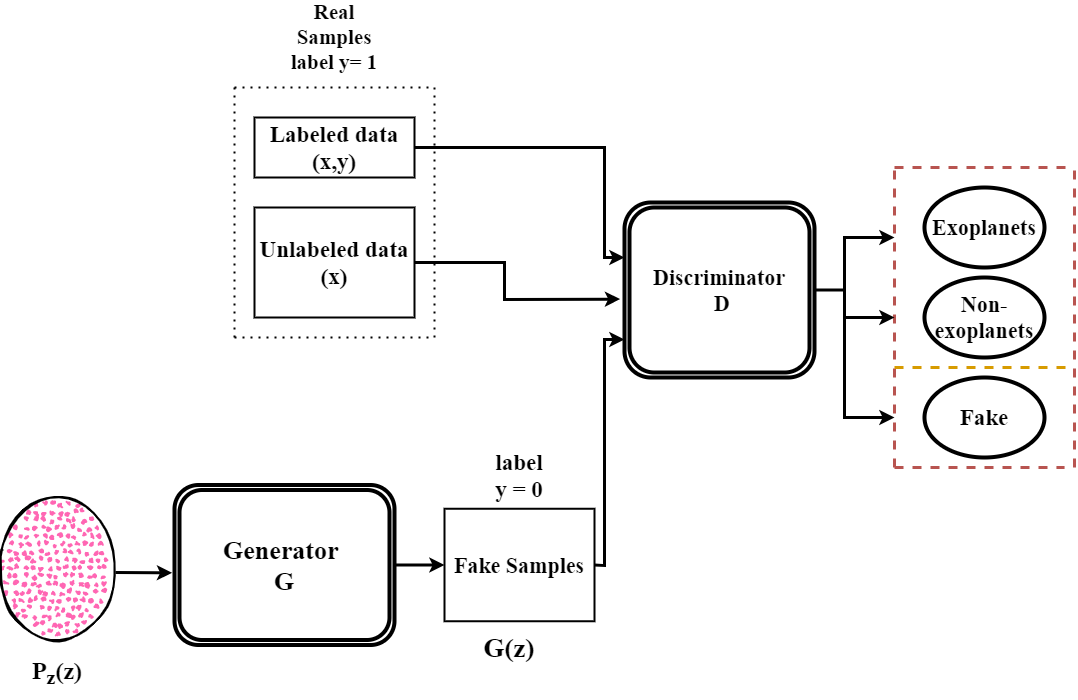}
\caption{Illustration of Architecture of ExoSGAN}
\label{fig:figure3}
\end{figure}
The unsupervised component has two parts. Given the data being real, one minus the expectation of the model producing the result as fake constitutes the first part. The next part is the expectation of the model producing the result as fake when the data is from G.
We can notice that substituting $D(x) = 1- P_{model}(y= N+1\vert x$) into the $L_{unsupervised}$ yields equation \ref{eqn2}.
\begin{equation}
\begin{aligned}
L_{unsupervised} \:= &-E_{x \sim P_{data}(x)} \log[1-P_{model}(y=N+1 \vert x)]\\
   & +E_{x \sim G} \log [P_{model}(y=N+1\vert x)]
\end{aligned}
\end{equation}
If the data is from any of the N classes, the supervised loss component is the negative log probability of the label.
\begin{equation}
L_{supervised} = -E_{x,y \sim P_{data}(x,y)}\log P_{model}(y\vert x,y<K+1)
\end{equation}
Therefore, the loss function  of SGAN can be given as
\begin{equation}
\begin{aligned}
L =& -E_{x,y \sim P_{data}(x,y)}[\log P_{model}(y \vert x)] \\
 & - E_{x \sim G[\log P_{model}}(y=N+1 \vert x)] 
\end{aligned}
\end{equation}

The output of the softmax does not change when a common function f(x) is subtracted from each of the classifier's output logits. So if we fix the logits of the N+1th class to 0, the value doesn’t change. 
Finally if $Z(x)  = \sum_{k=1}^{k} exp[lk(x)]$ is the normalized sum of the exponential outputs,
the discriminator can be given as
\begin{equation}
D(x)  = \frac{Z(x)}{Z(x)+1} 
\end{equation}

\subsection{Auxillary Classifier Generative Adversarial Networks(ACGAN)}

Mirza and Osindero \cite{mirza2014conditional} presented conditional GAN (CGAN), an enhancement to GAN, to conditionally generate samples of a specific class. The noise z  along with class labels c are given as input to generator of CGAN to generate fake samples of a specific class ($ X_{fake}=G(c,z) $). This variation improves the stability and speed while training GAN. The discriminator of CGAN is fed with class labels as one of the inputs. ACGAN is a CGAN extension in which instead of inputting class labels into D, class labels are predicted. It was introduced by Odena et al \cite{odena2017conditional}. The  discriminator of ACGAN can be thought of as having two classification functions: one that predicts whether the input is real or false(probability $ P(S\vert X) $), and another that classifies the samples into one of the classes(probability $ P(C\vert X) $). $ L_{C} $, the log-likelihood of the correct class and $ L_{S} $, the log likelihood of the source make up the two parts of the loss function of ACGAN.
\begin{equation}
L_{S}= E[\log P(S=real\vert X_{real})]+E[\log P(S=fake\vert X_{fake})]
\end{equation}  
\begin{equation}
L_{C}=E[\log P(C=c\vert X_{real})]+E[logP(C=c\vert X_{fake})]
\end{equation}
The generator and discriminator play a minmax game on $ L_{S} $ as in case of normal GAN. They try to maximise $ L_{c} $. In contrast to the conditional GAN, the resultant generator learns a latent space representation irrespective of the class label.
\begin{figure}[t]
\centering
\includegraphics[width=0.9\textwidth]{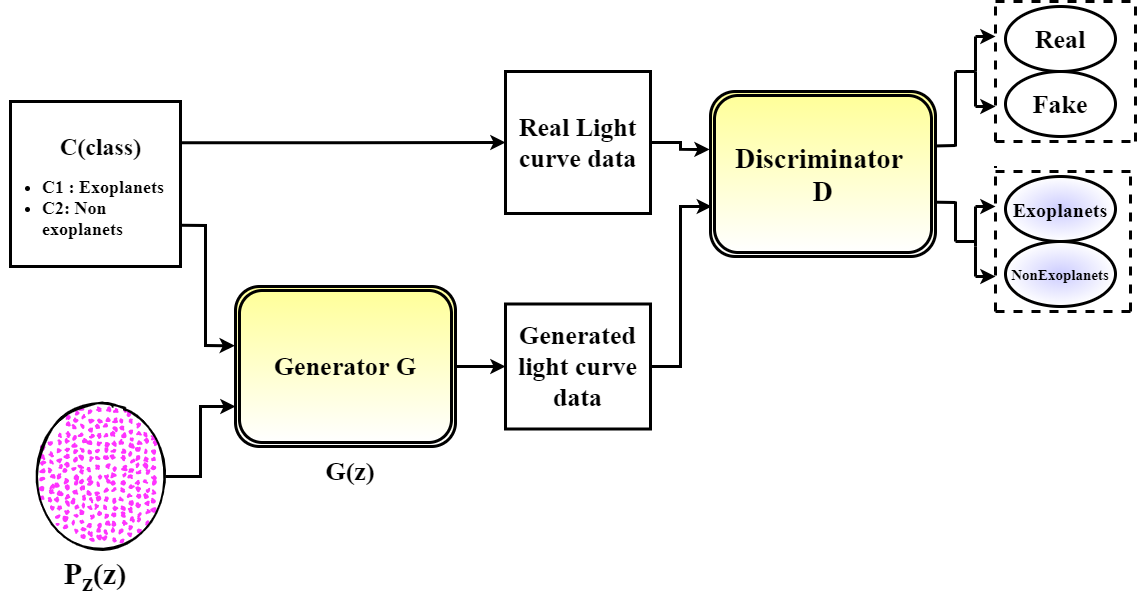}
\caption{Illustration of Architecture of ExoACGAN}
\label{fig:figure4}
\end{figure}
\section{Data Sources and Implementation}\label{sec4}
\subsection{Dataset}
\subsubsection*{Kepler Mission}
NASA began the Kepler mission on March 7, 2009, to study stars and hunt for terrestrial planets, particularly habitable planets with liquid water. Two of Kepler's four reaction wheels malfunctioned by May 11, 2013. At least three of them should be in good shape to keep the spaceship pointed in the right direction. Because the damage could not be repaired, the K2 -"Second Light"- was initiated (February 4, 2014), which takes advantage of Kepler's remaining capabilities. K2 conducted ecliptic-pointed ‘Campaigns’ of 80 days duration. The light intensities were recorded every 30 minutes. These stellar light intensities are used to look for dips in order to detect any probable exoplanets transiting the star. A total of 477 confirmed planets were found by K2 as of December 2021. We use data from K2's third campaign, which began on November 12th, 2014, for this study. A few samples from other campaigns were also used to enhance the number of stars containing exoplanets. However, Campaign 3 accounts for nearly all of the data. Almost 16,000 stars were included in the field-of-view(FOV) of the third Campaign. Kepler’s data was made public by NASA via the Mikulski Archive\footnote{\url{https://archive.stsci.edu/missions-and-data/kepler}}. We used the open-sourced data from Kaggle \cite{data} which was created by transposing the PDC-SAP-FLUX column(Pre-search Data Conditioning) of the original FITS file.
\begin{figure}[t]
\centering
\includegraphics[width=\textwidth,totalheight=0.35\textheight]{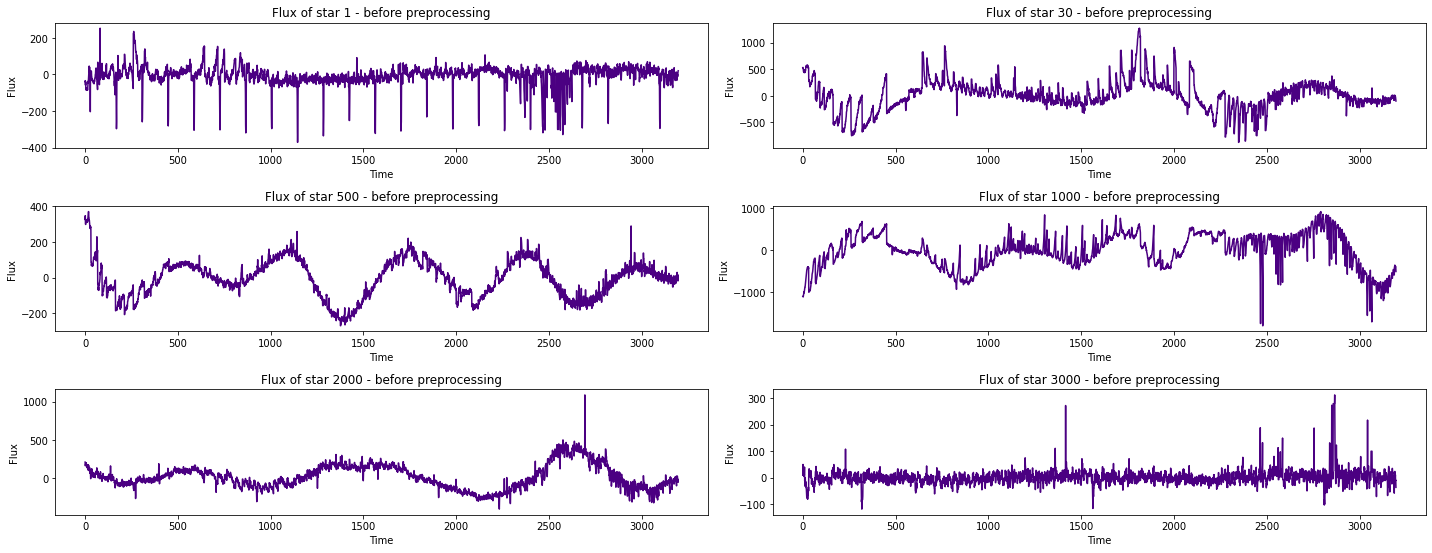}
\caption{Plot of the light curves of a few stars from the dataset with exoplanets and without exoplanets-before preprocessing. X-axis gives the time and Y-axis shows the flux intensities. The curves are not normalized and standardized. Also note the presence of huge outliers.}
\label{fig:figure5}
\end{figure}
\begin{figure}[t]
\centering
\includegraphics[width=\textwidth,totalheight=0.35\textheight]{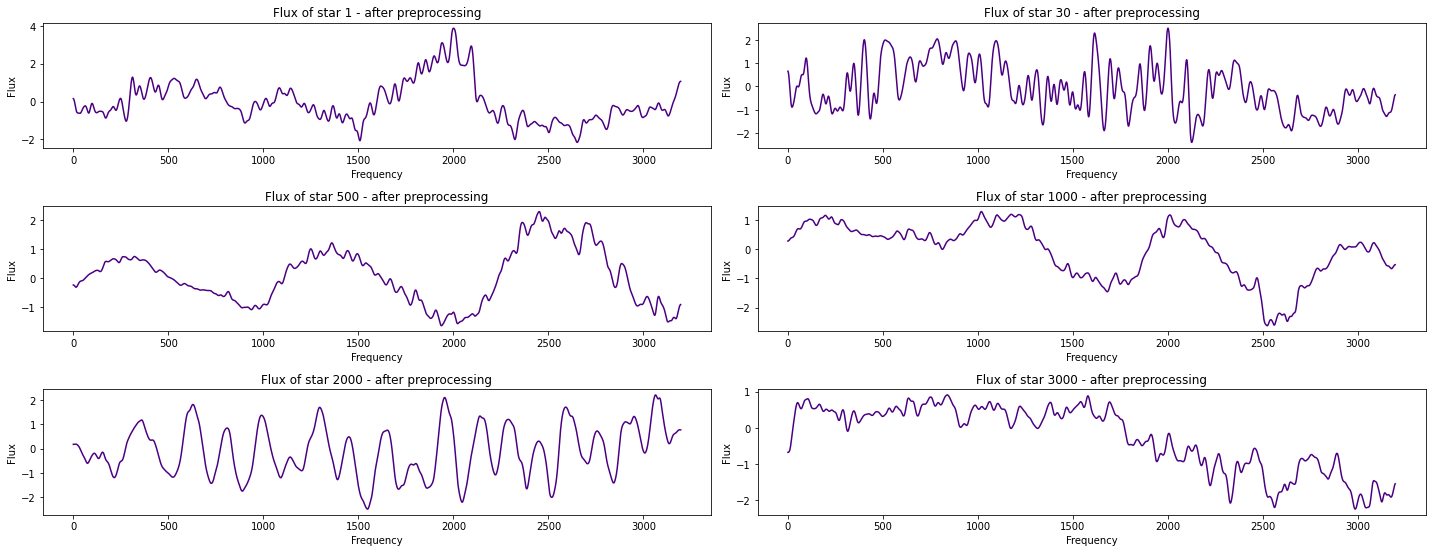}
\caption{Plot of the light curves of a few stars with exoplanets and without exoplanets-after preprocessing. X-axis shows the frequency and Y-axis plots the flux intensity. The curves are normalized and standardized as well as converted into frequency domain. The major upper outliers are also removed. }
\label{fig:figure6}
\end{figure}
\subsubsection*{Train and Test data: }

The train set includes 5087 observations, each with 3198 features. The class labels are listed in the first column. Label 2 denotes stars having exoplanets, whereas label 1 denotes stars without exoplanets. The same is true for the test set. The following columns 2-3198 show the intensity of light emitted by the stars at equal time intervals. As the Campaign lasted 80 days, the flux points were taken around 36 minutes apart (3197/80). There are 545 samples in the test data, 5 of which contain exoplanets. These are the unseen data used to test the method's outcomes. We have renamed the class labels, 1 for exoplanet-stars and 0 for nonexoplanet-stars for convenience. Figure \ref{fig:figure5} illustrates few samples from train data.

\subsection{Data Pre-processing}
\begin{enumerate}
\item The data contains no duplicates and null values as it has already been de-noised by NASA.
\item When we take a look at the data, we notice that there are many outliers in intensity values. Since we search for dips, it is important not to eliminate outlier clusters caused by transits that naturally belong to the target star.
So we remove a tiny fraction of the upper outliers and replace them with mean of adjacent points.
Later Fast Fourier Transform, followed by gaussian filter is applied to convert the raw flux from temporal to frequency domain and smoothen the curve. 
\item As the features are in various ranges, we should normalize the data such that the row values range between -1 and 1. We may also standardize the data by using StandardScaler to ensure the column values have a standard deviation of one. Figure \ref{fig:figure6} illustrates the light curves after the preprocessing steps. 
\end{enumerate}

\begin{table}[b]
    \centering
    \caption{Hyperparameters used in SGAN model}
    \label{table:table2}
    \begin{adjustbox}{width=\textwidth,totalheight=\textheight,keepaspectratio}
    \begin{tabular}[t]{c c c c c c c}
        \toprule
Operation & Kernel & Strides  &Feature Map & BN? &Dropout &Non-linearity \\ \midrule
Generator \\ \midrule
Dense       &N/A  &N/A   &800  &NO  &0.0  &NO\\
Transposed 1-d Convolution & $7\times 7$  & $2\times 2$  &128 &YES &0.0 &LeakyRelu\\
Dense       &N/A  &N/A   &1065  &YES  &0.0  &LeakyRelu\\
Dense       &N/A  &N/A   &1598  &YES  &0.0  &LeakyRelu\\
Dense       &N/A  &N/A   &3197  &NO  &0.0  &tanh\\ \midrule
Discriminator \\ \midrule

Transposed 1-d Convolution & $7\times 7$  & $1\times 1$  &32 &NO &0.0 &LeakyRelu\\
Max-Pooling 1d  & $2\times 2$ & $1\times 1$ &32 &NO &0.25 &- \\

Transposed 1-d Convolution & $7\times 7$  & $1\times 1$  &32 &NO &0.0 &LeakyRelu\\
Max-Pooling 1d & $2\times 2$ & $1\times 1$ &64 &NO &0.25 &- \\

Dense       &N/A  &N/A   &128  &NO &0.0  &LeakyRelu\\
Dense       &N/A  &N/A   &64  &NO &0.2  &LeakyRelu\\
Dense       &N/A  &N/A   &32  &NO &0.0  &LeakyRelu\\
Dense       &N/A  &N/A   &2  &NO  &0.0  &softmax\\
Lambda      &N/A  &N/A   &1   &NO  &0.0  &custom-activation\\ \midrule
Generator Optimizer &Adam $(\eta = 4e-6 , \beta_{1} = 0.5)$\\
Discriminator Optimizer &Adam $(\eta = 4e-5,  \beta_{1} = 0.5)$ \\
Batch size                          & 32\\
iteration                           & 3200\\
LeakyReLu slope                      & 0.1\\
Activation noise Standerd deviation & [0.2]\\ 

        \bottomrule
    \end{tabular}\hfill%
\end{adjustbox}
\end{table}

\subsection{Model Architecture and Training : ExoSGAN}
The generator G takes noise from a random normal distribution (standard deviation 0.02) in the latent space and generates fake flux curves, which are one of three inputs to D. The true dataset, which comprises the flux curve as well as its labels, is the next input to D. The final set of inputs comes from real samples as well, but this time they are unlabeled.

A convolutional neural network with two output layers, one with loss function binary cross entropy and the other with sparse categorical cross entropy, makes up the discriminator network. The second output layer uses a softmax activation function to predict the stars with and without exoplanets while the first output layer uses sigmoid activation to find the realness of the data. In the generator a transpose of 1 dimensional convolutional layer followed by dense layers is used. We also use BatchNormalisation and LeakyReLu with a slope of 0.2. A tanh activation function is employed in the output layer. 

We get a better performance model when we increase the number of labeled samples to 148 (74 positive and 74 negative samples). To slightly increase the positive labeled real samples we reverse the order of the flux points preserving the shape and add them to the original dataset. Now the train dataset contains 74 light curves with exoplanets and 5050 light curves with no exoplanets. Still the imbalance remains the same with ratio 1:100 exoplanet and non exoplanet stars. 

 The training procedure of GAN includes holding D constant while training G and vice versa. We use a learning rate $\eta=4e-6$ and $\beta_{1}= 0.5$. G is trained via D and Discriminator D is a stacked model with shared weights where the results can be reused.  The unsupervised model is stacked on top of the supervised model before the softmax activation function where the former takes output from the latter. 
Figure \ref{fig:figure3} shows the SGAN model architecture used in this paper and Table \ref{table:table2} gives the hyperparameters used in Exo-SGAN model.

\begin{table}[b]
    \centering
    \caption{Hyperparameters used in ACGAN model}
    \label{table:table3}
    \begin{adjustbox}{width=\textwidth,totalheight=\textheight,keepaspectratio}
    \begin{tabular}[t]{c c c c c c c}
        \toprule
Operation & Kernel & Strides  &Feature Map & BN? &Dropout &Non-linearity \\ \midrule
Generator \\ \midrule
Dense       &N/A  &N/A   &500  &NO  &0.0  &NO\\
Embedding   &N/A  &N/A   &10   &NO  &0.0  &NO\\
Dense       &N/A  &N/A   &500  &NO  &0.0  &Relu\\
Concatenate &N/A  &N/A   &N/A  &NO  &0.0  &NO \\
Transposed 1-d Convolution & $7\times 7$  & $2\times 2$  &128 &YES &0.0 &LeakyRelu\\
Dense       &N/A  &N/A   &1065  &YES &0.0  &LeakyRelu\\
Dense       &N/A  &N/A   &1598  &YES &0.0  &LeakyRelu\\
Dense       &N/A  &N/A   &3197  &NO  &0.0  &tanh\\ \midrule
Discriminator \\ \midrule
Transposed 1-d Convolution & $7\times 7$  & $2\times 2$  &32 &YES &0.0 &LeakyRelu\\
Transposed 1-d Convolution & $7\times 7$  & $2\times 2$  &64 &YES &0.0 &LeakyRelu\\
Transposed 1-d Convolution & $7\times 7$  & $2\times 2$  &128 &YES &0.0 &LeakyRelu\\
Transposed 1-d Convolution & $7\times 7$  & $2\times 2$  &128 &YES &0.0 &LeakyRelu\\
Transposed 1-d Convolution & $7\times 7$  & $2\times 2$  &128 &YES &0.8 &LeakyRelu\\
Dense       &N/A  &N/A   &1  &NO &0.0  &sigmoid\\
Dense       &N/A  &N/A   &2  &NO  &0.0  &softmax\\ \midrule
Generator Optimizer &Adam $(\eta = 4e-6 , \beta_{1} = 0.5 )$\\
Discriminator Optimizer &Adam $(\eta = 4e-5,  \beta_{1} = 0.5)$ \\
Batch size                          & 32\\
iteration                           & 3200\\
LeakyReLu slope                      & 0.2\\
Activation noise Standerd deviation & [0.2]\\ 

        \bottomrule
    \end{tabular}\hfill%
\end{adjustbox}
\end{table}

\subsection{Model Architecture and training: ExoACGAN}

In the architecture adopted here, we feed random noise and the class labels of exoplanet and non-exoplanet light curves into the generator, which subsequently produces synthetic data. The discriminator receives created flux intensity curves as well as real flux intensity curves as input. It guesses if the data is authentic or bogus and distinguishes between light curves with and without exoplanets. 

The discriminator is implemented as a 1-dimensional convolutional network with input shape(3197,1). The learning rate  of D is set to $ \eta=4e-5 $ while that of G is $ \eta = 4e-6 $. 
Similar to Exo-SGAN, D has two output layers where the first predicts real/fake class and the second gives the probability of the stars having exoplanets. The generator model takes in latent space with 100 dimensions and the single integer class labels. The latent vector is given to a dense layer and later reshaped. We use an embedding layer in G to feature map the class labels with a dimension of 10 (arbitrary). This can then be interpreted by a dense layer. Afterwards, the tensors formed by noise and class labels are concatenated producing an additional channel and sent through a 1-dimensional convolutional transpose layer, which is flattened and then passed through 3 dense layers later to finally produce curves of 3197 flux points.Also, LeakyRelu and Batch normalisation are used to provide regularization. The output layer is designed with a hyperbolic tangent activation function. ACGAN also uses a composite model like SGAN where the generator is trained via the discriminator. Figure \ref{fig:figure4} shows the SGAN model architecture used in this paper and Table \ref{table:table3} gives the hyperparameters used in Exo-SGAN model.

\section{Evaluation}\label{sec5}
\subsection{Metrics used for evaluation}

\begin{enumerate}

\item Confusion Matrix: The summarization of the classification performed by Exo-SGAN and Exo-ACGAN can be given by a confusion matrix. It provides information not only on the faults produced,
but also on the sorts of errors made. 
For binary classification, it is a 2x2 matrix of actual and predicted positive and negative classifications as given in Table \ref{table:table4}:
where, 

TP : True positive (A star with exoplanet predicted as a star with exoplanet)

FP: False positive (A star without exoplanet predicted as a star with exoplanet)

FN: False negative (A star without exoplanet predicted as star with exoplamet)

TN: True negative (A star without exoplanet predicted as a star without exoplanet) 
\begin{table}[b]

\caption{Confusion matrix for exoplanet classification}
\centering
\label{table:table4}
\begin{tabular}{cc|c|c|c|}
&\multicolumn{1}{c}{}&\multicolumn{3}{c}{\textbf{Predicted}}\\
&\multicolumn{1}{c}{}&\multicolumn{1}{c}{\textbf{Negative}}
&\multicolumn{1}{c}{\textbf{Positive}}
&\multicolumn{1}{c}{\textbf{Total}}\\

\cline{3-5}
\multicolumn{1}{c}{\multirow{3}{*}{\rotatebox{90}{\textbf{Actual}}}}
&\textbf{Negative} &TP &FP &TP+FP\\
\cline{3-5}
&\textbf{Positive} &FN &TN &FN+TN\\
\cline{3-5}
&\textbf{Total} &TP+FN &FP+TN &N\\
\cline{3-5}
\end{tabular}

\end{table}

\item Accuracy:  When evaluating a model for classification issues, accuracy is frequently used. As the name implies, accuracy is the number of right predictions divided by the total number of predictions. An accuracy of 1.00 means that all of the samples were properly categorized. However, in this situation, as the data we have is very imbalanced, accuracy alone cannot be used to evaluate the performance of the models since an 99 accuracy of 0.99 might also suggest that all of the samples are placed into the majority class. 
\begin{equation*}
Accuracy = \dfrac{TP+TN}{TP+FP+TN+FN}
\end{equation*}

\item
Precision: Precision seeks to address the issue of how many positive identifications were truly correct. A precision of 1 indicates that the model produced no FP. To find precision we use the formula
\begin{equation*}
Precision=\dfrac{TP}{TP+FP}. 
\end{equation*} 
  Precision is the ratio between the true positives to the total of true positives and false positives.
\item
Recall: The proportion of positive samples recovered is referred to as recall. It is a useful metric when we need to accurately categorise all of the positive samples. Sensitivity is another term for recall.
\begin{equation*}
Recall = \dfrac{TP}{TP+FN} 
\end{equation*} 
\item
Specificity: 
Specificity is found by dividing the true negatives by the total number of actual negative samples. Specificity answers the question of how many stars with no exoplanets did the model accurately predict.
\begin{equation*}
Specificity=\dfrac{TN}{TN+FP}
\end{equation*}
\item

F-beta score: The harmonic mean of precision and recall is used to get the F-score. F-beta is an abstraction of the F-score in which the balance of precision and recall in the computation of the harmonic mean is regulated by a coefficient called $\beta$. In the computation of the score, a beta value greater than 1.0, such as 2.0, provides more weight to recall and less weight to precision.
\begin{equation*}
F_{\beta} = \dfrac{(1+\beta^{2})(Precision \times recall)}{\beta^{2}Precision+Recall}
\end{equation*}
F-2 Score can be obtained by substituting $\beta = 2$
\begin{equation*}
F_{2} = \dfrac{5}{\dfrac{4}{Precision}+\dfrac{1}{Recall}}
\end{equation*}
\end{enumerate}

\begin{table}[b]
    \centering
   
    \caption{Confusion matrix of Exo-SGAN model on training dataset} 
    \label{table:table5}
{

\begin{tabular}{cc|cc|c}
\multicolumn{2}{c}{}
            &   \multicolumn{2}{c}{Predicted} \\
    &       &   without exoplanet &   with exoplanet     &Total          \\ 
    \cline{2-5}
\multirow{2}{*}{\rotatebox[origin=c]{90}{Actual}}
    & without exoplanet   & 5032   & 18    &5050             \\
    & with exoplanet    & 0    & 74   &74             \\ 
    \cline{2-5}
    & Total & 5032 & 92 & 5124 \\
    \cline{2-5}
    \end{tabular}
}

\end{table}

\begin{table}[t]
    \centering
    
    \caption{Confusion matrix of Exo-ACGAN model on training dataset} 
    \label{table:table6}
{

\begin{tabular}{cc|cc|c}
\multicolumn{2}{c}{}
            &   \multicolumn{2}{c}{Predicted} \\
    &       &   with exoplanet &   without exoplanet     &Total          \\ 
    \cline{2-5}
\multirow{2}{*}{\rotatebox[origin=c]{90}{Actual}}
    & with exoplanet   & 5041   & 9    &5050             \\
    & without exoplanet    & 0    & 74   &74             \\ 
    \cline{2-5}
    & Total & 5041 & 83 & 5124 \\
    \cline{2-5}
    \end{tabular}
}

\end{table}
\begin{table}[t]
    \centering
    
    \caption{Confusion matrix of Exo-SGAN and Exo-ACGAN model on testing data} 
    \label{table:table7}
{

\begin{tabular}{cc|cc|c}
\multicolumn{2}{c}{}
            &   \multicolumn{2}{c}{Predicted} \\
    &       &   without exoplanet &   with exoplanet     &Total          \\ 
    \cline{2-5}
\multirow{2}{*}{\rotatebox[origin=c]{90}{Actual}}
    & without exoplanet   & 565   & 0    &565             \\
    & with exoplanet    & 0    & 5   &5             \\ 
    \cline{2-5}
    & Total & 565 & 5 & 190 \\
    \cline{2-5}
    \end{tabular}
}

\end{table}

\subsection{Results and Discussion}
\begin{table}[t]
    \centering
    
    \caption{Results of the performance of Exo-SGAN model on train and test data}
    \label{table:table8}
    \begin{tabular}[t]{c c c}
        \toprule
        Exo-SGAN           & Training      & Testing      \\ \midrule
        Accuracy           & .996          & 100.0\\
        Precision          & 0.804         & 1.00 \\
       Recall              & 1.00          & 1.00 \\
        Specificity        & 0.996         & 1.00  \\
        F-score            & 0.954         & 1.00   \\
        \bottomrule
    \end{tabular}\hfill%

\end{table}
\begin{table}[t]
	
    \centering
    \caption{Results of the performance of Exo-SGAN model on train and test data}
    \label{table:table9}
    \begin{tabular}[t]{c c c}
        \toprule
        Exo-ACGAN           & Training      & Testing      \\ \midrule
        Accuracy           & .998          & 100.0\\
        Precision          & 0.892         & 1.00 \\
       Recall              & 1.00          & 1.00 \\
        Specificity        & 0.998         & 1.00  \\
        F-score            & .976            & 1.00   \\
        \bottomrule
    \end{tabular}\hfill%

\end{table}
The ExoSGAN and ExoACGAN models produce encouraging results in the classification of exoplanets. It should be noted that to the best of our knowledge semi-supervised generative adversarial networks and auxiliary classifier generative adversarial networks have not yet been used for the classification of exoplanet stars. Based on our findings, we believe that these approaches are extremely promising for detecting exoplanets from light curves. 

We argue that, among the assessment criteria described above, recall is the most important since we should not overlook any stars with exoplanets. As a result, we trained our models to provide the highest possible recall. Maximum recall is at the expense of precision. A model with high recall may have low precision. This is known as the precision-recall trade-off. In our case, having a few stars without exoplanets labelled as stars with exoplanets can be compromised if we can properly categorise each star with exoplanets. It is important to remember that the data is extremely biased with a majority of non exoplanet candidate stars. Therefore as mentioned earlier, accuracy cannot be considered as a proper metric for evaluation. As we place greater emphasis on recall, we increase the weight of recall in the F-score and set the beta value to 2.

The models, both ExoSGAN and ExoACGAN, output probabilities of a star containing exoplanet. The default threshold for such a classification task is often set at 0.5. Since we deal with imbalanced data and aim to take out all the lighcurves containing the transit, we should optimize the threshold. We choose our threshold to be 0.81 for the 
\begin{figure}[t]
\centering
\includegraphics[width=0.9\textwidth]{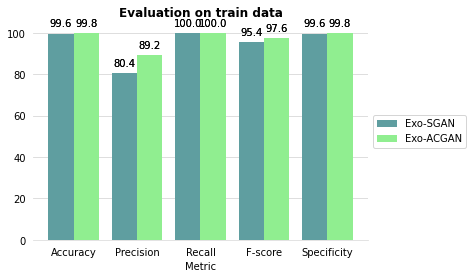}
\caption{A comparison of the outcomes of each of the Exo-SGAN and Exo-ACGAN assessment metrics using train data }
\label{fig:figure7}
\end{figure}
\begin{figure}[t]
\centering
\includegraphics[width=0.9\textwidth]{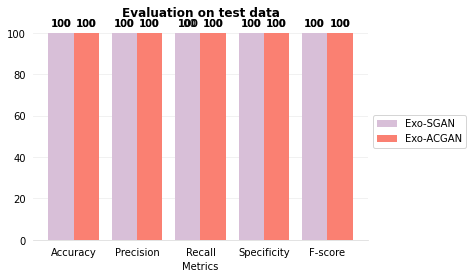}
\caption{A comparison of the outcomes of each of the Exo-SGAN and Exo-ACGAN assessment metrics using test data }
\label{fig:figure8}
\end{figure}
semi-supervised classification model and 0.77 for the auxiliary classifier generative adversarial network.
Both of these models give us a perfect recall of 1.00 on the train and test set for the chosen threshold. We get a precision of 0.802 from Exo-SGAN and 0.892 from Exo-ACGAN on the train data. While training, Exo-ACGAN results in an accuracy of 99.8  whereas Exo-SGAN gives an accuracy of 99.6. When we test our models on unseen data both Exo-ACGAN and Exo-SGAN give a perfect accuracy of 100 percent. All the 565 non-exoplanet stars and 10 exoplanet stars in the test set are classified correctly. Therefore, the precision, recall, specificity, and f-score turns out to be 1.00 for both Exo-SGAN and Exo-ACGAN on test data as shown in Table \ref{table:table8} and \ref{table:table9}. On the train samples, out of 5050 non-exoplanet stars only 8 have been misclassified by Exo-ACGAN, and only 18 have been misclassified by Exo-SGAN. The confusion matrix in Table \ref{table:table5}, \ref{table:table6} and \ref{table:table7}  tabulates the classification results of our experiments.

When we compare Exo-SGAN and Exo-ACGAN, we can find that ACGAN marginally surpasses Exo-SGAN. Both have a recall of one, but the latter has a gain of 0.2, 0.09, and 0.09 in accuracy, precision, and specificity while training.

It can be speculated that during the training and testing procedure of Exo-ACGAN, the adversarial training technique assisted in reducing the impact of the imbalanced dataset. As a result, the prejudice towards the majority class is reduced. The discriminator model is taught to learn patterns from both actual and generated data in an adversarial manner. As a consequence, in order to generalize the patterns from the training dataset, the discriminator model learnt a greater number of richer patterns.

\section{Conclusions and Future works}\label{sec6}
In this paper, we utilise the Semi-Supervised Generative Adversarial framework and Auxilary Classifier Generative Adversarial framework to detect exoplanets using the flux curves of stars. The performance of both methods is noteworthy since both of them properly detect all lightcurves with transit. Exo-SGAN and Exo-ACGAN both has a recall rate of 100. The capacity to use unlabeled candidates to get better outcomes is the key advantage of our proposed network Exo-SGAN. We anticipate that, as the number of exoplanet candidates increase and maintaining a big labelled dataset becomes more difficult, this method will become even more beneficial for future exoplanet detection. Even though the data is extremely unbalanced, Exo-ACGAN and Exo-SGAN are able to produce better results even without the use of any upsampling approaches like SMOTE.

As for the next steps, these architectures can be tried out on the raw lightcurves from the Kepler mission, provided Graphical Processing Units(GPU) and hardware of great performance since a large amount of data must be handled. The hyperparameters can also be tuned to give better performance. Additionally, experimenting with various data-preprocessing approaches for eliminating outliers may improve the performance.  One method to attempt in order to solve the class imbalance issue is to optimize the loss function to produce data only from the minority class while training \citep{8392662}. This can bring a balance between the samples of both classes. Yet another technique to improve the classification accuracy is the use of Bad GAN \citep{dai2017good} in which the generator’s goal is more focused to produce data that complements the discriminator’s data.

However, our models can provide a very reliable system for finding all the true positives in our data at the current stage. It is hoped that these deep learning approaches would pave the way for a new era in the field of exoplanet hunting.

\section*{Declarations}
We declare that this manuscript is original, has not been published before,
and is not currently being considered for publication elsewhere.
\begin{itemize}
\item Funding: No funding was received for conducting this study. 
\item Conflict of interest: The authors declare no conflict of interest.
\item Ethics approval: NA 
\item Consent to participate: NA
\item Authors' contributions: Conceptualization: Cicy K Agnes; Methodology: Akthar Naveed V; Formal analysis and investigation: Cicy K Agnes, Akthar Naveed V; Project adimistration and Validation: Cicy K Agnes; Supervision: Anitha Mary M O Chacko; Writing - original draft preparation: Cicy K Agnes; Writing - review and editing: Anitha Mary M O Chacko, Akthar Naveed V;

\end{itemize}

\bibliography{sn-bibliography}

\end{document}